\documentclass[twocolumn,preprintnumbers,showkeys]{revtex4}

\usepackage[T1]{fontenc}
\usepackage{graphicx}

\usepackage{amsmath}
\usepackage{amssymb}
\usepackage{amsthm}
\usepackage{amsfonts}
\usepackage{enumerate}
\usepackage{bm}

\usepackage{centernot}
\usepackage{tikz}

\usepackage{subfigure}

\begin{document}

\title{Solving local constraint condition problem in slave particle theory with the BRST quantization}

 \author{Xi Luo$^1$}
 \author{Jianqiao Liu$^2$}
\author{ Yue Yu$^{2}$}
\thanks{Correspondence to: yuyue@fudan.edu.cn}
 \affiliation {  1. College of Science, University of Shanghai for Science and Technology, Shanghai 200093, China\\
 2.Department of Physics, Fudan University, Shanghai 200433,China\\}
 

\begin{abstract}  With the Becchi-Rouet-Stora-Tyutin (BRST) quantization of gauge theory, we solve the long-standing difficult problem of the local constraint conditions, i.e., the single occupation of a slave particle per site, in the slave particle theory. This difficulty is actually caused by inconsistently dealing with the local Lagrange multiplier $\lambda_i$ which ensures the constraint: In the Hamiltonian formalism of the theory,  $\lambda_i$ is time-independent and commutes with the Hamiltonian while in the Lagrangian formalism, $\lambda_i(t)$ becomes time-dependent and plays a role of gauge field.  This implies that the redundant degrees of freedom of $\lambda_i(t)$ are introduced and must be removed by the additional  constraint, the gauge fixing condition $\partial_t \lambda_i(t)=0$. In literature, this gauge fixing condition was missed.  We add this gauge fixing condition and use the BRST quantization of gauge theory for Dirac's first-class constraints in the slave particle theory. This gauge fixing condition endows $\lambda_i(t)$ with dynamics and leads to important physical results. As an example, we study the Hubbard model at half-filling and find that the spinon is gapped in the weak $U$ and the system is indeed a conventional metal, which resolves the paradox that the weak coupling state is a superconductor in the previous slave boson mean field theory. For the $t$-$J$ model, we find that the dynamic effect of $\lambda_i(t)$ substantially suppresses the $d$-wave pairing gap and then the superconducting critical temperature may be lowered at least a factor of one-fifth of the mean field value which is of the order of 1000 K.  The renormalized $T_c$ is then close to that in cuprates. 
\end{abstract} 

\maketitle
\section{introductions}

The Hubbard model, though it is simple, is in the central position for understanding  strongly correlated electron systems \cite{Hubbard1,Hubbard2}.The single-band Hubbard model \cite{Anderson} was considered as the starting point to explain the high-$T_c$ superconductivity (SC) \cite{highTCSC}.  The strong Hubbard repulsion limit of the Hubbard model tends to the $t$-$J$ model, which also was derived from a more realistic model for cuprates \cite{tJ1,tJ2,ZR}.  Numerous subsequent studies on these two models were done either analytically or numerically.   Many numerical simulation results are very impressive but they are basically subject to the computational resources and so are far from  conclusive ones.  Useful analytical approaches include the Gutzwiller approximation \cite{Gutz},  mean field (MF) theories \cite{B, Ru,zou, affl,SF,RMFT,RGU1,RGU2,RGU3}, and the gauge theory \cite{LN,WL,WNL,Lee, MuF1,MuF2,Sach} based on the slave particle formalism \cite{Bar1,Bar2,Col,KR,KL,SHF,LWH}.
Analog to the slave particle models,  a large class of models, e.g., spin-fermion models, were developed to study the strongly correlated systems such as cuprates based on the spin fluctuations \cite{s-f1,s-f2,s-f3,s-f4,weng}.

Recently, the renormalized MF theory based on the Gutzwiller projection \cite{RMFT,RGU1,RGU2,RGU3} has been generalized to that in the form of statistically-consistent Gutzwiller approximation \cite{SGA}, which was proved to be equivalent to the slave boson theory.  The results in terms  of the further subsequent generalization,  i.e., a systematic diagrammatic expansion of the variational Gutzwiller-type wave function may be quantitatively compared with the experimental properties of  cuprates \cite{PRSGA}.

We are not going to focus on the results obtained by these methods  because they are too fruitful to be summarized. We will try to improve the slave boson method and fix some shortcomings of the theory. For example, the slave particle theory looks very powerful because it exactly maps a strongly correlated electron system to a weakly coupled slave particle one but the things become difficult when dealing with local constraint conditions $T_i=0$ (see Eq. (\ref{lc})), i.e.,  only one type of the single slave particle can occupy a lattice site $i$.  The temporal component of the gauge field, $\lambda_i(t)$ and the spatial components of the gauge field which are introduced to compensate the gauge symmetry breaking by the MF approximation are not dynamic so that the conventional perturbation theory is not applicable. In this paper, we try to  solve these problems.

In terms of Dirac's  approach to solve the first-class constraint systems,  a term $-\sum_i\lambda_iT_i$ with $\lambda_i$ being the Lagrange multiplier is added to the Hamiltonian $H$. Since there are no temporal or spatial derivatives of $\lambda_i$ in the Hamiltonian, $[H,\lambda_i]=0$ and then $\lambda_i$ will not evolve with time. In literature, $\lambda_i$ was simply relaxed to a time-dependent field $\lambda_i(t)$ and as the temporal component of the gauge field. This introduces the redundant degrees of freedom because $\lambda_i$ should be kept static, i.e.,  an additional constraint $\partial_t\lambda_i(t)\equiv\dot\lambda_i(t)=0$ must be enforced. This point was missed before. Instead, in the MF approximation, a conventional approximation $\lambda_i(t)=\bar\lambda$, a constant with no spatial and temporal dependence, was taken.  Although $\dot {\bar\lambda}$ is zero, obviously this brings many unphysical degrees of freedom so that the MF theory after this approximation is not reliable or controllable. Many further improvements are proposed to deal with this issue but they do not bring conclusive results  \cite{LN,WL,WNL,Lee}. Recent development in the statistical Gutzwiller approximation sheds light to systematically relieving the difficulties that original renormalization mean field theory meets \cite{SGA,PRSGA}. In this paper, we make efforts to improve the slave particle theory by considering the additional constraint $\dot\lambda_i(t)=0$ instead of $\lambda_i=\bar\lambda$. 

When an electron operator is decomposed into slave particles, a gauge symmetry is induced and $\lambda_i(t)$ behaves as a gauge potential in the temporal direction.  To remove the redundant gauge degrees of freedom, one has to introduce a gauge fixing condition (GFC) while keeping the physical observables is gauge invariant. Simply setting $\lambda_i(t)=\bar\lambda$ is a GFC but it is not a good GFC because it violates the constraint $T_i=0$ and brings unphysical degrees of freedom. For a gauge theory with constraints,  the GFC must be consistent with the constraints.  

In this paper, we show $\dot \lambda_i(t)=0$ is a good GFC because it keeps  $T_i=0$ unchanging.  Introducing this GFC may be thought of as an application of the general method of the gauge fixing in the gauge theory with Dirac's first-class constraints \cite{FV1,FV2}.  After gauge fixing, the gauge invariance of the system is equivalent to the Becchi-Rouet-Stora-Tyutin (BRST) global symmetry.  For Dirac's first-class constraints,  {\it the BRST symmetry is the criterion whether the GFC is correct} because the BRST invariance requirement to the physical states is exactly equivalent to Dirac's first-class constraints. 

 The constraint $\dot\lambda_i(t)=0 $ endows  $\lambda_i(t)$ with the dynamics. The fluctuation from $\lambda_i(t)$ through the interaction between $\lambda_i(t)$ and the slave particles enables us to examine the stability of  the variational wave function of a given phase.  In this sense, the variational or MF states of the phases become controllable and we then solve the local constraint condition problem of the slave particle theory.  For different purposes, we may deal with $\lambda_i(t)$ in different ways. In the gauge theory of the slave boson, in order to study the normal state behavior of the systems, the spatial components of the gauge field are introduced  \cite{LN,WL,WNL,Lee}. The physical properties of the system were studied by integrating away the matter fields. The spatial components of the gauge field actually also play the role of the Lagrange multiplier of the counterflow constraint on the slave particles' currents. The similar additional constraint to $\dot\lambda_i(t)=0$ is also needed but was not considered in the previous researches. In this paper, we will not consider the spatial components of the gauge fields because there might be further complicated symmetry analysis more than the gauge symmetry in that case. We will leave it for a coming soon work.  
 
In this paper, we focus on the pairing states of the spinons and study the BCS-type MF states. In this case, the spatial components of the gauge field are not necessary to be introduced.   We examine the spinion pairing gap of the Hubbard and $t$-$J$ models. For the Hubbard model, it was found that the MF state of the slave bosons at half-filling in a small $U$ is an $s$-wave SC state \cite{wen1,wen2}. This is obviously wrong because the Hubbard model in the weak coupling limit is a conventional metal. We show that after considering the dynamics of $\lambda_i(t)$ induced by the constraint $\dot\lambda_i(t)=0$, the SC state is not stable because integrating over $\lambda_i(t)$ induces an unusual pairing instability of spinon's Fermi surface and the spinon is gapped. This destroys the SC at half-filling.  

Similarly, for the $t$-$J$ model at the spinon pairing gap state or the SC state, integrating over $\lambda_i(t)$ contributes an additional unusual term to the spinon pairing. This additional contribution does not destroy the MF SC gap but may substantially suppress it. Numerically, the gap will be smaller than at least a factor of  one-fifth of the MF SC gap. Thus, one can expect  the $d$-wave SC critical temperature $T_c$, whose MF value is of the order 1000 K,  is substantially lowered and might be comparable with that of cuprates. 
 
This paper is organized as follows:  In Sec. II,  we explain why the additional constraint $\dot\lambda_i(t)=0$ is necessary for the slave particle theory. We relate this additional constraint to the BRST quantization of the gauge theory. In Sec. III, we apply our theory to the Hubbard model at half-filling and show the SC state in the small $U$ is unstable. In Sec. IV,  we study the spinon gap suppression by $\lambda_i(t)$'s fluctuation in the $t$-$J$ model and discuss the implication to $T_c$ of curpates. In Sec. V, we conclude this work and schematically look forward to the prospect of the applications of our theory to the strongly correlated systems.  
 
 \section{Constraint to the Lagrange multiplier and BRST quantization}  
 
 For a strongly correlated electron many-body system, a conventional perturbation theory based on the Fermi liquid theory does not work.  In order to turn the strong interacting electron model to an equivalent weak coupling theory, a powerful method called the slave boson/fermion theory is applied \cite{Bar1,Bar2,Col,KR}. For the electron operator $c_{i\sigma}$ at a lattice site $i$, the local quantum space is $\{|0\rangle, |\uparrow\rangle,|\downarrow\rangle, |\uparrow\downarrow\rangle\}$. The completeness condition reads
 \begin{eqnarray}
 |0\rangle\langle0|+|\uparrow\rangle\langle\uparrow|+|\downarrow\rangle\langle\downarrow|+|\uparrow\downarrow\rangle\langle\downarrow\uparrow|=1. \label{compness}
 \end{eqnarray} 
 The slave boson representation of the electron is mapping $|0\rangle\to h^\dag, |\sigma\rangle\to f^\dag_\sigma$ and $|\uparrow\downarrow\rangle\to d^\dag$. The operators $h,d,f_\sigma$ are called  the holon, doublon, and spinon which destroy some vacuum $|vac\rangle$.  For the slave boson, $h$ and $d$ are the bosonic operators while $f_\sigma$ are the fermionic operators
 \begin{eqnarray}
 [d_i,d^\dag_j]=[h_i,h^\dag_j]=\delta_{ij}, \{f_{i\sigma},f^\dag_{j\sigma'}\}=\delta_{ij}\delta_{\sigma,\sigma'},
 \end{eqnarray}
  and so on.  The completeness condition (\ref{compness}) maps to a local constraint
 \begin{eqnarray}
 T_i=h^\dag_i h_i+f^\dag_{i\uparrow} f_{i\uparrow}+f^\dag_{i\downarrow} f_{i\downarrow}+d^\dag_i d_i-1=0,\label{lc}
 \end{eqnarray}
 i.e.,  a given lattice site can only be occupied by one given type of particles.   
 The electron operator is decomposed into $c^\dag_{i\sigma}= f^\dag_{i\sigma}h_i +\sigma f_{i,-\sigma}d_i^\dag$ with $\sigma=\{\uparrow,\downarrow\}\equiv\{+,-\}$. With the local constraint, the anti-commutation of the electron operators are equivalent to $h,d$ s' are bosonic while $f_\sigma$ s' are fermionic. This equivalence also holds if $f_\sigma$ are bosonic and $h,d$ are fermionic, which is called the slave fermion representation. In this work, we focus on the slave boson one although they are equivalent before  
MF approximations. Using the slave boson representation, a strongly correlated many-body electron Hamiltonian $H_e$ can be mapped to a Hamiltonian $H$ in the slave boson representation.  

\subsection{Additional Constraint }

For a Hamilton system with constraints, we follow Dirac's method to solve the constrained system and introduce a Lagrange multiplier $\lambda_i$. The Hamiltonian for the constrained problem is given by
\begin{eqnarray}
H_\lambda=H-\sum_i\lambda_iT_i. \label{LH}
\end{eqnarray}
In Schr\"odinger's picture, $H$, $\lambda_i$ and $T_i$ are all time-independent.  Notice that $[H_\lambda,\lambda_i]=0$ and then $\lambda_i$ does not evolve as time.  

Going to Heisenberg's picture, all operators and fields $\Phi_i$ become time-dependent,  $\Phi_i(t)=\mathrm{e}^{\mathrm{i}H_\lambda t}\Phi_i\mathrm{e}^{-\mathrm{i}H_\lambda t}$ except $\lambda_i$ since $[H_\lambda,\lambda_i]=0$, where  $ \Phi_i$ stand for the `matter' fields (holon, doublon, and spinon). On the other hand,  because the constraint $T_i=0$ has to be enforced in any space-time location, the Lagrange multiplier has to be relaxed to time-dependent. Thus, one has to add an additional constraint  in order to be consistent with no time evolution of  $\lambda_i(t)$, namely,
\begin{eqnarray}
\dot\lambda_i(t)=0.
\end{eqnarray}
By introducing a new Lagrange multiplier $\pi_{\lambda_ i}(t)$ to force $\dot \lambda_i(t)=0$, the Lagrangian is then given by 
\begin{eqnarray}
L_\lambda&=&\sum_i\pi_{\lambda_i}(t)\dot\lambda_i(t)+\sum_{i\sigma}f^\dag_{i\sigma}(i\partial_t+\lambda_i(t)) f_{i\sigma}\nonumber\\
&+&\sum_i(h^\dag_i(i\partial_t+\lambda_i(t))h_i+\sum_id^\dag_i(i\partial_t+\lambda_i(t))d_i\nonumber\\
&-&\sum_i\lambda_i(t)-H \label{La}.
\end{eqnarray}

Another way to understand the relation between the Hamiltonian (\ref{LH}) and the Lagrangian (\ref{La}) is as follows.  According to (\ref{La}), $\pi_{\lambda i}(t)=\frac{\delta L}{\delta \dot\lambda_i(t)}$, i.e., $\pi_{\lambda i}(t)$ is the canonical conjugate field of $\lambda_i(t)$.  Therefore, according to the classical mechanics, the Lagrangian of the Hamiltonian (\ref{LH}) reads
\begin{eqnarray}
L_\lambda=\sum_i (\pi_{\lambda i} \dot \lambda_i+\Pi_{\Phi_i} \dot \Phi_i)-H_\lambda, \label{Laa}
\end{eqnarray}
where $\Pi_{\Phi_i}$  are the canonical conjugate fields of $ \Phi_i$. The Lagrangian (\ref{Laa}) is exactly the same as (\ref{La}).

\subsection{Gauge Symmetry}

 We now explain the reason to add the constraint $\dot\lambda_i(t)=0$ from the gauge symmetry point of view. 
 In literature,  instead of (\ref{La}), the following Lagrangian is considered  \cite{B,zou}  
 \begin{eqnarray}
 L_{\mathrm{GI}}&=&\sum_{i \sigma}f^\dag_{i\sigma}(\mathrm{i}\partial_t+\lambda_i(t)) f_{i\sigma}+\sum_i(h^\dag_i(\mathrm{i}\partial_t+\lambda_i(t))h_i
 \nonumber\\
 &+&\sum_id^\dag_i(\mathrm{i}\partial_t+\lambda_i(t))d_i-\sum_i \lambda_i (t) -H \label{LGI}.
\end{eqnarray}
It was known that the electron operator $c^\dag_{i\sigma}= f^\dag_{i\sigma}h_i +\sigma f_{i,-\sigma}d_i^\dag$ is gauge invariant under  $(h_i,d_i, f_{i\sigma})\to \mathrm{e}^{-\mathrm{i} \theta_i} (h_i,d_i, f_{i\sigma})$. $L_{\mathrm{GI}}$ is invariant under this gauge transformation accompanying  with $\lambda_i(t)\to \lambda_i(t)-\dot\theta_i$, i.e., $\lambda_i(t)$  plays a role of a scalar gauge potential.  There are redundant gauge degrees of freedom in the path integral 
\begin{eqnarray}  
 W'&=&\int \prod_{i,t} \mathrm{d}\Phi_i^\dag(t) \mathrm{d}\Phi_i(t) \mathrm{d}\lambda_i(t)\mathrm{e}^{\mathrm{i}\int \mathrm{d} t L_{\mathrm{GI}} }.\label{W'}
\end{eqnarray}
One way to remove the redundant gauge degrees of freedom is replacing $L_{\mathrm{GI}}$ by the Lagrangian (\ref{La}), 
\begin{eqnarray}
W=\int \prod_{i,t} \mathrm{d}\Phi_i^\dag(t) \mathrm{d}\Phi_i(t) \mathrm{d}\pi_{\lambda_i}(t) \mathrm{d}\lambda_i(t)\mathrm{e}^{\mathrm{i}\int \mathrm{d}t L_\lambda }. \label{W}
\end{eqnarray}
 Making a transformation $\dot\lambda_i\to\dot\lambda_i+\xi \pi_{\lambda i}/2$ for Eq. (\ref{W}) where $\xi$ is an arbitrary constant and integrating away $\pi_{\lambda i}$ field, the path integral reads
 \begin{eqnarray}
W\propto \int \prod_{i,t} \mathrm{d}\Phi_i^\dag(t) \mathrm{d}\Phi_i(t) \mathrm{d}\lambda_i(t)  \mathrm{e}^{\mathrm{i}\int \mathrm{d}t L_{\rm eff}},\label{W1}
 \end{eqnarray}
where 
\begin{eqnarray}
L_{\rm eff}=L_{\mathrm{GI}}-\frac{1}{2\xi}\sum_i\dot\lambda_i^2(t).
\end{eqnarray}
This is a correct gauge fixing Lagrangian of the Abelian gauge theory but Eq. (\ref{W1}) is not gauge invariant.
In order to resolve this paradox,  we recall Faddeev-Popov quantization of the gauge theory. We insert 1 into the gauge invariant (\ref{W'}) to fix the redundant gauge degrees of freedom in terms of
\begin{eqnarray}
1=\int \prod_{i,t}\mathrm{d} \theta_{i,t} \delta(\dot\lambda_i(t)){\rm det}(\frac{\delta\dot\lambda_i(t)}{\delta\theta_j(t')}),
\end{eqnarray}
and finally \cite{Peskin}
\begin{eqnarray}
1\cdot W'&=&N(\xi)\int \prod_{i,t} \mathrm{d}\Phi_i^\dag(t) \mathrm{d}\Phi_i(t) \mathrm{d}\lambda_i(t){\rm det }(\partial^2_t)\nonumber\\
&\times&\exp\{\mathrm{i}\int \mathrm{d}t L_{\rm eff}\},\label{GFL}
\end{eqnarray}
where $N(\xi)$ is an unimportant infinity constant. The path integral (\ref{GFL}) is gauge invariant.  Comparing (\ref{GFL}) and (\ref{W1}), they differ from a factor ${\rm det }(\partial^2_t)$ after dropping $N(\xi)$. At the present case, this determinant does not contain any fields and is a constant. This means that  (\ref{W1}) is equivalent to (\ref{GFL}). Therefore, up to a constant determinant, (\ref{W1})  is gauge invariant.  However, for a non-Abelian gauge theory, the determinant in general is dependent on the gauge field and can not be dropped. This is why Faddeev-Popov ghost fields are introduced.  

\subsection  {BRST Quantization} 

Historically, there is a standard approach to deal with the relativistic gauge theory with Dirac's first-class constraints $T^\alpha ({\bm r},t)=0$ with  $[T^\alpha,T^\beta]=f^{\alpha\beta}
 _\gamma T^\gamma$  for constants $f^{\alpha\beta}
 _\gamma$ \cite{FV1,FV2}. It is called the BRST quantization of a gauge theory which is the generalization of the Faddeev-Popov path integral quantization of a gauge theory.  Fradkin and Vilkovisky pointed out that in such relativistic gauge theories, it is necessary to include the time derivative of all Lagrange multiplier fields to be the GFCs, i.e., $\partial_t\lambda_\alpha({\bf r},t)+F_\alpha(\pi_{\lambda _\alpha})=0$ with an arbitrary function $F_\alpha(\pi_{\lambda_\alpha})$ \cite{FV1}. {\it For Dirac's first-class constrained systems, the BRST symmetry gives a criterion whether the GFC is correct.}  Applying their approach to the present case with an Abelian constraint $T_i(t)=0$ and taking $F(\pi_{\lambda i})=\xi \pi_{\lambda i}$, the BRST invariant Lagrangian is given by
 \begin{eqnarray}
 L_{\rm BRST}=L_{\rm eff}+\sum_i\bar u_i\partial_t^2 u_i, \label{BRST}
 \end{eqnarray} 
 where $u_i$ and $\bar u_i$ are the Faddeev-Popov ghost and anti-ghost fields, the fermionic fields obeying $\{u_i,\bar u_j\}=\delta_{ij}$. It is easy to check $L_{\rm BRST}$ is invariant under the BRST transformations 
  \begin{eqnarray}
\delta_Bu_i=0,\delta_B \bar u_i=\epsilon \dot\lambda_i/\xi,  \delta_B\lambda_i=-\epsilon \dot u_i,
\end{eqnarray}
where $\epsilon$ is an anti-commuting constant with $\epsilon^2=0$ while the slave particles' transformations are  $h_i,d_i, f_{i\sigma}$ varying a local phase $\mathrm{e}^{-\mathrm{i}\epsilon u_i(t)}=1-\mathrm{i}\epsilon u_i(t)$. 
It is easy to check that $\delta_B^2=0$. This is called the nilpotency of the BRST transformation. 
The BRST quantized path integral is given by
\begin{eqnarray}
 W_{\rm BRST}=\int \prod_{i,t} \mathrm{d}\Phi_i^\dag(t) \mathrm{d}\Phi_i(t) \mathrm{d}\lambda_i(t) \mathrm{d} \bar u_i(t) \mathrm{d}u_i(t) \mathrm{e}^{\mathrm{i}\int \mathrm{d}t L_{\rm BRST}}.\nonumber
\end{eqnarray}
Integrating over the ghost fields, the path integral $W_{\rm BRST}$ recovers the path integral (\ref{W1}). Therefore, the BRST quantization of the gauge theory is exactly equivalent to the conventional path integral quantization. Notice that this BRST quantization may also be applied to non-Abelian gauge theory such as the SU(2) gauge theory of the slave boson \cite{WNL}.  In the quantization of the non-Abelian gauge theory, the determinant for the non-Abelian gauge theory will not be easily treated without introducing the Faddeev-Popov ghosts.     

The benefits gained from the BRST quantization are that :

(1) Because the BRST symmetry is a global symmetry with respect to the fermionic constant $\epsilon$, one can define the conservation fermionic charge from Nother's theorem
\begin{eqnarray}
Q=-\sum_i(u_{i}T_i+\frac1{\xi}\dot\lambda_i\dot u)
\end{eqnarray}
All the physical states which are ghost-free obey
\begin{eqnarray}
Q|{\rm Phys}\rangle=0.\label{cocycle}
\end{eqnarray}
This recovers $T_i=0$ and $\dot\lambda_i(t)=0$. From the gauge theory point of view,  
$\lambda_i(t)=\bar\lambda$ is also a GFC, which removes the redundant degrees of freedom but the constraint $T_i=0$ is relaxed to $\langle T_i\rangle=0$. This brings other unphysical degrees of freedom into the quantum state space. The gauge theory developed in \cite{LN,WL,WNL,Lee} tried to solve this problem in a different way from us. 

(2) The nilpotency $Q^2=0$ resembles the external differential operator $d^2=0$ in the deRahm cohomology. The constraint (\ref{cocycle})  is called a BRST cocycle condition and all physical states form the BRST cohomology group which topologically classifies the strongly correlated systems.

(3) Introducing the ghost fields greatly simplifies the quantization of  the non-Abelian gauge theory which we will not involve in here. For the Abelian gauge theory considered in this paper, the ghost fields are decoupled to the gauge field and can be integrated away. Therefore, we will use the path integral (\ref{W1}). For finite temperature $T$, if mapping $t\to i\tau$, the path integral turns to the partition function
\begin{eqnarray}
Z=\int \prod_{i,\tau} \mathrm{d}\Phi_i^\dag(\tau) \mathrm{d}\Phi_i(\tau) \mathrm{d}\lambda_i(\tau)  \mathrm{e}^{-\int_0^\beta \mathrm{d}\tau L_{\rm eff}},
\end{eqnarray}
where $\beta=1/T$ and $\dot\lambda_i(\tau)\equiv \partial_\tau\lambda(\tau)$.

\section{The Hubbard model at half-filling }

To be concrete, we take the repulsive Hubbard model on a square lattice at half-filling as an example. The model Hamiltonian is given by 
\begin{eqnarray}
H=-t\sum_{\langle ij\rangle,\sigma}c^\dag_{i\sigma}c_{j\sigma}+U\sum_in_{i\uparrow}n_{i\downarrow},
\end{eqnarray}
where the hopping $t$ is fixed in between nearest neighboring
sites.  $U$ is the on-site Hubbard repulsion and $n_{i\sigma}=c^\dag_{i\sigma}c_{i\sigma}$.
In the slave boson representation, the Hamiltonian reads 
\begin{eqnarray}
H_\lambda&=&-t\sum_{\langle ij\rangle}[\chi^f_{ij}\chi^b_{ji}+\Delta^{f\dag}_{ij}\Delta^b_{ij}+\mathrm{h.c.}]+U\sum_id^\dag_id_i
\nonumber\\&-&\sum_i\lambda_iT_i
-\mu\sum_i( d^\dag_i d_i-h^\dag_ih_i),
\end{eqnarray}
where $\chi_{ij}^f=\sum_\sigma f_{i\sigma}^\dag f_{j\sigma}$, $\chi^b_{ij}=h^\dag_ih_j-d_i^\dag d_j$, $\Delta^f_{ij}=\sum_\sigma \sigma f_{i,-\sigma}f_{j\sigma}$, and $\Delta_{ij}^b=d_ih_j+h_id_j$. 
  
 As we have argued, to study the detailed properties the various phases, we need to do various MF approximation fluctuated by the spatial components of the gauge field which is not the task in this work. We only restrict on the fluctuation from $\lambda_i(t)$ and examine the instability of the BCS MF states at half-filling for small $U$. The original Hubbard model at half-filling is metallic for small $U$, while the previous slave boson MF theory gave a SC phase \cite{wen1,wen2}. In this SC phase, charge and spin excitations are gapless and the slave bosons condense \cite{wen1}.  Neglecting the boson fluctuation of the condensate, the effective Lagrangian reads
\begin{eqnarray}
L^s_{\rm eff}&=&\sum_{{\bm k} \sigma}f^\dag_{{\bm k} \sigma}\omega f_{{\bm k}\sigma}-\frac{t\chi^b}2\sum_{{\bm k}\sigma}  k^2 f^\dag _{{\bm k}\sigma} f_{{\bm k}\sigma}\nonumber\\
&-& \sum_i(\frac1{2\xi}\dot \lambda^2_i-\lambda_i n_{fi}),
 \end{eqnarray}
 where $t\chi^b=2t\rho_h=m_0^{-1}$ with $\rho_h$  the holon density and $n_{fi}=\sum_\sigma f^\dag_{i\sigma} f_{i\sigma}$.   We take
\begin{eqnarray}
 \lambda_i=\bar\lambda+ga_i, 
 \end{eqnarray}
 where $g$ is a constant which is arbitrary according to the constraint $T_i=0$.  The $a$-dependent part in $L^s_{\rm eff}$ reads
 \begin{eqnarray}
L_{a,f}=g\sum_{\bm{kq}\sigma} a_{\bm k}f^\dag_{\bm k+\bm q\sigma}f_{{\bf q}\sigma}-\frac{g^2}{2\xi}\sum_{{\bm k}}\dot a_{\bm k}\dot a_{-\bm k},~
 \end{eqnarray}
 where the first term is the interaction between the fluctuation field $a_{\bm k}=\sum_i a_i \mathrm{e}^{-\mathrm{i}{\bm k}\cdot {\bm r}_i}$ and the spinon.  
 The perturbation calculation for a small $g$ finds that the renormalization constant $Z_F$ does not vanish at spinon Fermi surface and then the gapless spinon liquid is stable against the perturbation fluctuation.
  
\begin{figure}
	\includegraphics [width=1.0\linewidth]{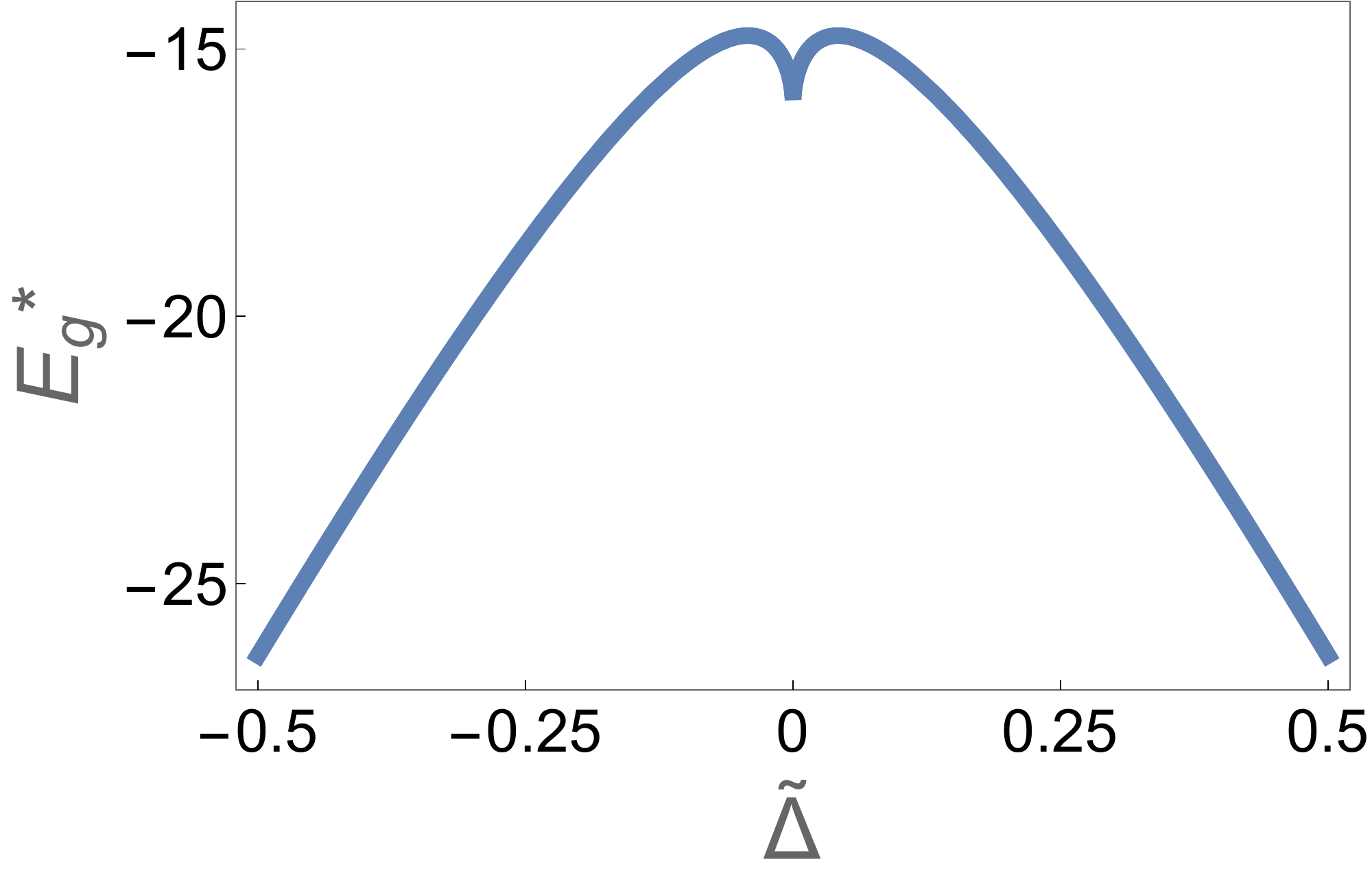}
	\caption{(color online)
		Numerical result of $E_g^*$ of $\tilde{\Delta}$. We choose the dispersion of the normal state to be $\xi_{\bm k}=-t(\cos k_x+\cos k_y)-\mu_f$, with $t=0.5$ and $\mu_f=-0.01$.} 
	\label{fig1}	
\end{figure}

 To see the instability, we rewrite $L_{a,f}$, 
\begin{eqnarray}
&&L_{a,f}(\omega)=g\sum_{\bm k \bm q\nu\sigma} a_{\bm k}(\omega)f^\dag_{\bm k+\bm q \sigma}(\omega+\nu)f_{{\bf q}\sigma}(\nu)\nonumber\\
&&-\omega^2\frac{g^2}{2\xi}\sum_{{\bm k}} a_{\bm k}(\omega)a_{-\bm k}(-\omega). \label{laf}
\end{eqnarray}
Integrating away $a_{\bm k}(\omega)$, the effective interacting Lagrangian between the spinons reads
\begin{eqnarray}
L^s_{\rm int}(\omega)&=&\frac{\xi}{2\omega^2}\sum_{\bm k \bm q\nu\sigma} f^\dag_{-{\bm k-\bm q},\sigma}(-\omega+\nu) f_{-{\bm k},\sigma}(\nu)\nonumber\\
&\times&\sum_{{\bm q}'\nu'\sigma'} f^\dag_{{\bm k+\bm q'},\sigma'}(\omega+\nu')f_{{\bm k}\sigma'}(\nu').
\end{eqnarray}
The pairing Hamiltonian is then given by
\begin{eqnarray}
H_{\rm pair}(\omega)&=&-\frac{\xi}{2\omega^2}\sum_{{\bm k \bm q}\nu\sigma}f^\dag_{{-\bm k-\bm q},\sigma}(-\omega-\nu)
 f^\dag_{{\bm k+\bm q},-\sigma}(\omega+\nu)\nonumber\\
 &\times&f_{-{\bm k},-\sigma}(-\nu)f_{{\bm k},\sigma}(\nu). 
\end{eqnarray}
For the pairing parameter $\tilde\Delta=\xi\langle f_{-{\bm k},\uparrow}f_{{\bm k},\downarrow}\rangle$, the MF energy is given by
\begin{eqnarray}
\omega^2-\frac{4\tilde\Delta^2}{\omega^4}-(\frac{{\bm k}^2}{2m_0}-\bar\lambda)^2=0. \label{disp}
\end{eqnarray}
When $\tilde\Delta=0$, the spinon is gapless at spin Fermi surface. When $\tilde\Delta\ne0$, the spinon has a  pairing gap is $\Delta^*=\pm\tilde\Delta^{1/3}$ and the spinon liquid is gapped. In this case, the spinon Fermi surface is not stable. Nor is the SC state of the Hubbard model at half-filling. It is metallic state because the charge excitation is still gapless.  To see the paired spinon liquid has a lower ground state energy than that of the gapless spinon, we consider the ground state energy 
\begin{eqnarray}
E_g=\sum_{\bm k}(-\frac{2\xi_{\bm k}^2+4\tilde\Delta^2/\omega^2}{2\omega})+\cdots\equiv E^*_g+\cdots,
\end{eqnarray}
where  $\xi_{\bm k}=\frac{{\bm k}^2}{2m_0}-\bar\lambda$ and $\cdots$ is $\tilde\Delta$-independent while $\omega$ can be solved from the dispersion relation (\ref{disp}) 
\begin{widetext}
 \begin{equation}
	\omega^2=\frac{1}{3}\left( \xi_{\bm k}^2+\frac{\xi_{\bm k}^4}{\sqrt[3]{\xi_{\bm k}^6+54\tilde\Delta^2+6\sqrt{3}|\tilde\Delta|\sqrt{\xi_{\bm k}^6+27\tilde\Delta^2}}} 
	+\sqrt[3]{\xi_{\bm k}^6+54\tilde\Delta^2+6\sqrt{3}|\tilde\Delta|\sqrt{\xi_{\bm k}^6+27\tilde\Delta^2}} \right).
\end{equation}
\end{widetext}
In Fig. \ref{fig1}, we plot a typical curve of the $\tilde{\Delta}$-dependent part of $E_g$, i.e., $E_g^*$ versus $\tilde{\Delta}$.  Here the lattice value of $\xi_{\bm k}$ is used and the summation of ${\bm k}$ runs over the first Brillouin zone. $\bar\lambda$ is converted to the spinon chemical potential $\mu_f$.  We see that the larger $\tilde \Delta$ is, the lower $E^*_g$ is.
On the other hand, in the MF theory, the gap $\tilde\Delta$ is determined by the self-consistent equation
\begin{equation}
	\tilde\Delta=\frac{\xi}{\omega^2}\sum_{\bm k}\frac{\tilde\Delta}{\sqrt{\xi_{\bm k}^2+(\frac{2\tilde\Delta}{\omega^2})^2}}.
\end{equation}
In the large $\tilde \Delta$ limit, $\omega^2\approx2.314\tilde \Delta^{2/3}$ and the self-consistent equation is approximated by
\begin{eqnarray}
	\sum_{\bm k}\frac{\xi}{2\tilde\Delta}\approx 1,
\end{eqnarray}
 i.e., the $\xi$-independent gap is $\Delta =\tilde\Delta/\xi\approx N_e/2$. Thus, the spinon is gapped. This means that the $s$-wave SC at half-filling for small $U$ is unstable and the system is a conventional metal. 

\section{$t$-$J$ model}  

We are going to the large $U$ limit.  It was known that the SC MF theory of the $t$-$J$ model has a SC $T_c\sim 1000$ K. Let us see if $\lambda_i$ fluctuation can suppress it. The $t$-$J$ model Hamiltonian on the square lattice is given by
\begin{eqnarray}
H_{\rm{t-J}}=-t\sum_{\langle ij\rangle \sigma}c^\dag_{i\sigma}c_{j\sigma}+J\sum_{\langle ij\rangle \sigma}({\bm S}_i\cdot{\bm S}_j-\frac{1}4 n_in_j),
\end{eqnarray}
where $S^a_i=\frac{1}2 \sum_{\sigma,\sigma'}c^\dag_{i\sigma}\sigma^a_{\sigma\sigma'}c_{i\sigma'}$ are the spin operators and 
$\sigma^a$ ($a=x,y,z$) are Pauli matrices.  The constraint is that there is no double occupation on each lattice site. The slave boson decomposition is now $c^\dag_{i\sigma}= f^\dag_{i\sigma}h_i$ and the local constraints are 
\begin{eqnarray}
T^{\rm{tJ}}_i=h_i^\dag h_i+\sum_\sigma f_{i\sigma}^\dag f_{i\sigma}-1=0.
\end{eqnarray}
 The effective slave boson $t$-$J$ Lagrangian then reads 
\begin{eqnarray}
L^{\rm{tJ}}_{\rm eff}&=&\frac{J}4 \sum_{\langle ij\rangle}[|\chi^f_{ij}|^2+|\Delta^f_{ij}|^2-(\chi^{f\dag}_{ij}\sum_\sigma f^\dag_{i\sigma} f_{j\sigma}+\rm h.c.)]\nonumber\\
&+&\sum_{\langle ij\rangle}\frac{J}4[\Delta^f_{ij}(f_{i\uparrow}^\dag f_{j\downarrow}^\dag-f_{i\downarrow}^\dag f_{j\uparrow}^\dag)+\rm {h.c.}]\nonumber\\
&+&\sum_i[h^\dag_i(i\partial_t-\mu)h_i+\mu x]+\sum_{i\sigma}f^\dag_{i\sigma}\mathrm{i}\partial_t f_{i\sigma}\nonumber\\
&-&\sum_i(\frac1{2\xi}\dot \lambda^2_i-\lambda_iT^{tJ}_i)-t\sum_{\langle ij\rangle}h_ih^\dag_jf^\dag_{i\sigma}f_{j\sigma}.\label{t-J}
\end{eqnarray}
Replacing $L_{\rm eff}$ in Eq. (\ref{W1}) by $L^{tJ}_{\rm eff}$, one has the path integral for the $t$-$J$ model. If we ignore the condensed holon fluctuation, the effective  low-lying Lagrangian in the SC MF state is given by 
\begin{eqnarray}
\bar L_{\rm eff}^{\rm tJ-sc}&=&\sum_{{\bm k} \sigma}f^\dag_{{\bm k} \sigma}\omega f_{{\bm k}\sigma}-\sum_{{\bm k}\sigma }\xi_{\bm k}f^\dag _{{\bm k}\sigma} f_{{\bm k}\sigma}\nonumber\\
&-&\sum_{\bm k}(\Delta_k f_{{\bm k}\uparrow}^\dag f_{-{\bm k}\downarrow}^\dag+h.c.)+L_{a,f},
\end{eqnarray}
where $\xi_{\bm k}=-(J\chi^f/4+t\rho_h)(\cos k_x+\cos k_y) -\mu_f$ and $\Delta_{\bm k}=J(\Delta_x \cos k_x+\Delta_y\cos k_y)$ with $\mu_f $ being the  chemical potential of the spinon and $\rho_h$ is the holon density. $L_{a,f}$ is given by (\ref{laf}).  The holon is condensed and the SC gap function is $\rho_h\Delta_k$.   Similarly, after integral over $a$-field,  there is an additional pairing term, $H_{\rm pair}$ in the MF Hamiltonian. The MF dispersion is then given by
\begin{eqnarray}
\omega^2-\Delta_k^2(1+\frac{2\xi}{J\omega^2})^2-\xi_{\bm k}^2=0. \label{tjd}
\end{eqnarray}
At the Fermi surface, the energy gap $\omega(\bm{k_F})=\Delta^*_{\bm{k_F}}$ is determined by 
\begin{eqnarray}
\hat \Delta_F^6-(\hat\Delta_F^2+A_F)^2=0, \label{tjd1}
\end{eqnarray}
where $\hat\Delta_F=\frac{\Delta^*_{\bm k_F}}{\Delta_{\bm k_F}}$ and $A_F=\frac{2\xi}{J\Delta_{\bm k_F}^2}$. 

{\it Because the MF approximation breaks the  gauge symmetry, the calculation results are dependent on the gauge fixing parameter $\xi$.}  We treat $\xi$ as a phenomenological parameter and so is $A_F(\xi)$. Obviously, Eq.(\ref{tjd1}) has solutions in $0\leq \hat\Delta_F\leq 1$ for $\xi\ne 0$ and $\hat\Delta_F=0$ or 1 only if $\xi=0$.   This means that the SC MF phase is stable but the gap parameter $\Delta^*_{k_F}$ is suppressed, e.g.,  when $A_F\sim -10^{-2}\pm10^{-3}$, $y\sim 0.1$. The renormalized pairing gap may be suppressed by one order from the MF gap $\Delta_{k_F}$.  This may explain why  $T_c\propto \rho_h\Delta^*_{k_F} $ in cuprates is one order lower than the MF estimation. 

\begin{figure}
	\includegraphics[width=1.0\linewidth]{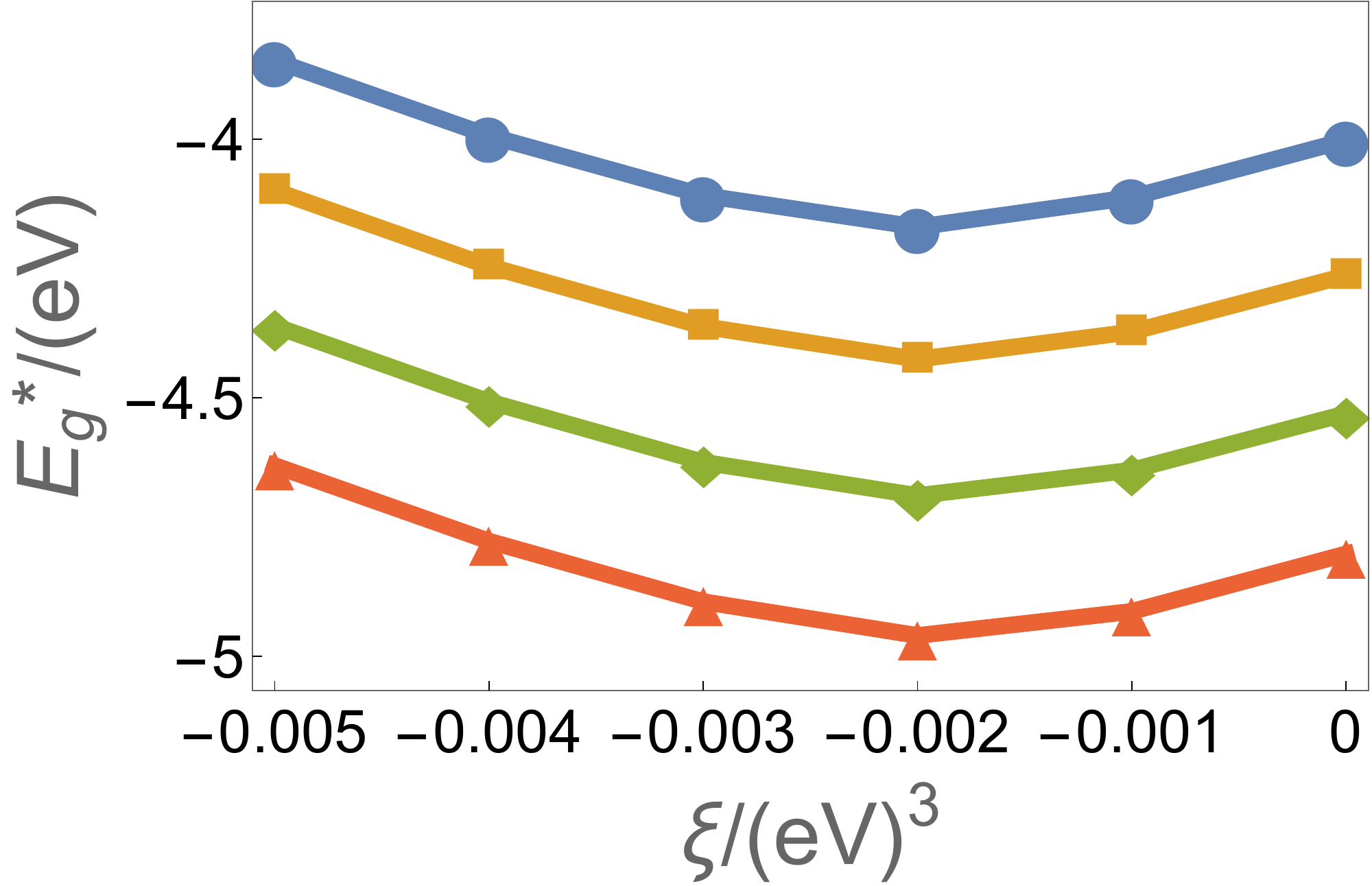}
	\caption{(color online)
		Numerical result of $E_{\rm g}^*$ of $\xi$. We choose the material data of cuprates \cite{cup}, i.e., $J\sim 0.12$ eV, $\chi^f\sim 0.2-0.3$, $\rho_h\sim 0.18-0.25$, and $\mu_f\sim -0.05$ eV. The four curves in the diagram, from the top to bottom, correspond to $J\chi^f/4+t\rho_h$ = 0.10 eV, 0.11 eV, 0.12 eV, and 0.13 eV, respectively.} 
	\label{fig2}	
\end{figure}

We can directly solve the dispersion relation (\ref{tjd}) and find that  
\begin{eqnarray}
\omega=\omega_{\bm k}=\sqrt{\xi^2_{\bm k}+\Delta^{*2}_{\bm k}},
\end{eqnarray}
where $\Delta^*_k$ is the renormalized gap (See Appendix A). The ground state energy of the BCS state is given by 
\begin{eqnarray}
E_{\rm{g}}(\xi)&=&\frac{1}{2J}\sum_{\bm k}\Delta^{*2}_k(\xi)-\sum_{\bm k}\omega_{\bm k}(\xi)+\cdots \nonumber \\
&\equiv& E^*_{\rm g}+\cdots,~
\end{eqnarray}
where $\cdots$ is $\xi$-independent terms.  With the material data of cuprates, we plot the $\xi$-dependent part of the $d$-wave SC MF ground state energies  $E^*_{\rm g}$ varying as $\xi$ in Fig. \ref{fig2}.  We find that $\xi\sim -0.002$ minimizes $E^*_{\rm g}$. With this value of $\xi$, we have $\frac{\Delta^*_{\bm k}}{\Delta_{\bm k}}\sim 0.2$  for $k_x=0,  k_y=\pi$. This lowers the MF $T_c$ a factor of one-fifth,  i.e., from $T^{\rm MF}_c\sim 1000$ K to $T^*_c\sim 200 $ K  (for more details,  see Appendix A).   

\section{Conclusions and prospects}

We have properly dealt with the local constraint conditions in the slave boson representation of the strongly correlated systems.   We argued that as a gauge theory with Dirac' s first-class constraint, taking the gauge fixing condition that removes the redundant gauge degrees of freedom must be consistent with the constraint. The BRST quantization is a consistent method to do that although the final path integral for the Abelian gauge theory is decoupled to the ghost fields.  We have applied our theory to the Hubbard model at half-filling and found that the ground state of the system in small $U$ is indeed a conventional metal. We showed that the MF s-wave SC state obtained by the slave boson representation in a previous study is not stable against the gauge fluctuation of the gauge field $\lambda_i(t)$. For the strong coupling system, we studied the $t$-$J$ model. We focused on the gauge fluctuation to the $d$-wave SC gap and found that it was substantially suppressed to a factor of one-fifth. For cuprates, this means that the MF SC $T_c$ is lowered from 1000 K to 200 K.  As we have mentioned, the gauge fluctuation from the spatial components of the gauge field was not considered. It might further reduce $T_c$ to be comparable to that of the cuprates materials. 
 
Historically,  the MF phase diagram of the $t$-$J$ model was studied in early days when the high $T_c$ SC was found in cuprates. The gauge fluctuation to various MF states was studied. However, as we see here, the GFC might not be introduced properly because the spatial components of the gauge field also play a role of the Lagrange multiplier to the constraint on the currents.  Additional constraints are also needed. This may endow the gauge field with dynamics. Then, the gauge invariant physical quantities can be calculated with perturbation theory. For instance,  one can calculate the renormalized pairing gap by Dyson's equation using the perturbation theory
\begin{eqnarray}
\frac{\Delta^*_{\bm k}(\omega)}{\Delta_{\bm k}}=\frac{G^{-1}_{\uparrow\downarrow}({\bm k},\omega)}{G^{-1}_{0\uparrow\downarrow}({\bm k})}=1-\frac{\Sigma_{\uparrow\downarrow}({\bm k},\omega)}{\Delta_{\bm k}},
\end{eqnarray}
where $G_{\uparrow\downarrow}$ is the anomalous spinon Green's function and $G_{0\uparrow\downarrow}^{-1}=\Delta_k$.
$\Sigma_{\uparrow\downarrow}$ is the anomalous self-energy of the spinon. This also implies $\Delta_k$ is suppressed to $\Delta^*_k(\omega)$ by the fluctuation. However, to establish a complete gauge theory with all components of the gauge fields, many symmetry considerations must be taken care of. We leave them to the further works.

This work is in memory of Professor Zhong-Yuan Zhu for his discussions with YY in the possible application of the BRST quantization to strongly correlated systems  thirty years ago. The authors thank Professor Qian Niu for his insightful comments and resultful discussions. We are grateful to Jianhui Dai and Long Liang for useful discussions.  This work is supported by NNSF of China with No. 12174067.

\appendix

\section{Dispersion and Renormalized Pairing Gap}

 The SC dispersion relation in the $t$-$J$ model can be solved by Eq. (\ref{tjd})
\begin{eqnarray}
	\omega^2-\Delta_{\bm k}^2(1+\frac{2\xi}{J\omega^2})^2-\xi_{\bm k}^2=0.\label{a1}
\end{eqnarray}
Defining
$$\hat E_{\bm k}^2=\frac{\omega^2}{\Delta_{\bm k}^2}, ~~\hat\xi_{\bm k}^2=\frac{\xi_{\bm k}^2}{\Delta_{\bm k}^2},A_k=-\frac{2\xi}{J\Delta^2_{\bm k}}, $$
Eq. (\ref{a1}) reads
 \begin{equation}
	\hat{E}_{\bm k}^2-(1-\frac{A_{\bm k}}{\hat{E}_{\bm k}^2})^2-\hat{\xi}_{\bm k}^2=0.
\end{equation}
For $E_{\bm k}^2>0$, we have 
\begin{eqnarray}
	\hat E^6-(\hat\xi_{\bm k}^2+1)\hat E_{\bm k}^4+2A_k\hat E_{\bm k}^2-A_{\bm k}^2=0.
\end{eqnarray}
Define $y=E_{\bm k}^2-(\hat\xi_{\bm k}^2+1)/3$, the above equation becomes,
\begin{equation}
	y^3+py+q=0, \label{3}
\end{equation}
with
\begin{eqnarray}
	p&=&\frac{6A_{\bm k}-(\hat\xi^2_{\bm k}+1)^2}{3},\nonumber\\
	q&=&\frac{-2(\hat \xi^2_{\bm k}+1)^3+18(\hat \xi^2_{\bm k}+1)A_{\bm k}-27A_{\bm k}^2}{27}.\nonumber
\end{eqnarray}
Then
\begin{eqnarray}
	{\cal D}&=&\frac{q^2}4+\frac{p^3}{27}\nonumber\\
	&=&\frac{1}{108} A_{\bm k}^2 \left(27 A_{\bm k}^2-4 A_{\bm k} (9 \xi_{\bm k}^2+1)+4 \xi_{\bm k}^2  (\xi_{\bm k}^2 +1)^2\right).\nonumber
\end{eqnarray}
It is easy to see that for a small $A_k$,  up to $\mathcal{O}(A_k^2)$, ${\cal D}>0$. Therefore, the solution of Eq. (\ref{3}) is 
\begin{equation}
	y=\sqrt[3]{-\frac{q}{2}+\sqrt{(\frac{q}{2})^2+(\frac{p}{3})^3}}+\sqrt[3]{-\frac{q}{2}-\sqrt{(\frac{q}{2})^2+(\frac{p}{3})^3}}.
\end{equation}
Then
\begin{equation}
	\hat E^2_{\bm k}=y+(\hat\xi_{\bm k}^2+1)/3.
\end{equation}
This gives the dispersion relation. 
Defining the renormalized gap by
$$\Delta_{\bm k}^{*2}=E^2_{\bm k}-\xi^2_{\bm k},$$ 
we have 
\begin{eqnarray}
\Delta_{\bm k}^{*2}=\Delta_{\bm k}^{2}(y-\frac{1}3(2\hat\xi_{\bm k}^2-1)).
\end{eqnarray}
The ground state energy is given by
	\begin{equation}
		E_g=\frac{1}{2J}\sum_{\bm k} \Delta_{\bm k}^{*2}-\sum_{\bm k}\omega_{\bm k}+\cdots\equiv E_g^*+\cdots,
\end{equation}
where the $\cdots$ stands for the $\xi$-independent part. In the numerical calculation, we choose the parameters of cuprates, i.e., $J\sim 0.12$ eV, $\chi^f\sim 0.2-0.3$, $\rho_h\sim 0.18-0.25$, and  $J\chi^f/4+t\rho_h\sim 0.1-0.13$ eV \cite{cup}. The spinon chemical potential $\mu_f$ is determined by 
	\begin{equation}
		1-\rho_h=\rho_f=-\frac{1}{N}\frac{\partial G}{\partial \mu_f},
	\end{equation}
where $\rho_f$ is the spinon density and $G$ is the Gibbs free energy. To the zeroth order of $\xi$, it reduces to 
	\begin{equation}
		\rho_f=\frac{1}{N}\sum_{\bm k} (1-\frac{\xi_{\bm k}}{\omega_{\bm k}}\tanh (\frac{\omega_{\bm k}}{2k_{\mathrm{B}}T})),
	\end{equation}
where $N$ is the number of total particles. By varying $T$ from 200 K to 1000 K, $\mu_f$ keeps $\sim -0.05$ eV. From the numerical results (see Fig. \ref{fig2} in the main text), $\xi=-0.002$ minimizes the ground state energy. At $k_x=0$, $k_y=\pi$ and $\Delta_k\sim 0.17$ eV,  $\Delta_k^*\sim 0.0318$ eV. Therefore, the critical temperature becomes $T_c\sim 1000\frac{\Delta_{\bm k}^*}{\Delta_{\bm k}}\sim 200$ K.


\end{document}